\documentclass{revtex4}

\def\bea{\begin{eqnarray}}
\def\eea{\end{eqnarray}}
\def\bec{\begin{center}}
\def\ec{\end{center}}

\def\beq{\begin{equation}}

\def\f{\frac}

\def\f#1#2{\frac{#1}{#2}}

\def\pr{\prime}
\def\l{\left}
\def\r{\right}

\def\bea{\begin{eqnarray}}
\def\eea{\end{eqnarray}}
\def\bec{\begin{center}}
\def\ec{\end{center}}

\def\f#1#2{\frac{#1}{#2}}

\def\l{\left}
\def\r{\right}

\def\f{\frac}

\def\beq{\begin{equation}}
\def\eeq{\end{equation}}

\def\beq{\begin{equation}}
\def\eeq{\end{equation}}

\begin{document}
\draft
\title{Gauge Unification and Flavor Hierarchy
from Extra 
Dimensions\footnote{\uppercase{T}alk presented 
at {\it \uppercase{SUSY} 2003:
\uppercase{S}upersymmetry in the \uppercase{D}esert}\/, 
held at the \uppercase{U}niversity of \uppercase{A}rizona,
\uppercase{T}ucson, \uppercase{AZ}, \uppercase{J}une 5-10, 2003.
\uppercase{T}o appear in the \uppercase{P}roceedings.}}
\author{Kiwoon Choi\footnote{kchoi@hep.kaist.ac.kr}}
\affiliation{Department of Physics, Korea Advanced Institute of 
Science and Technology\\ Daejeon
305-701, Korea}
\date{\today}
\begin{abstract}
Extra dimension provides an attractive way to break
symmetry by boundary conditions which can be useful to
construct a natural grand unified theory
avoiding the doublet-triplet
splitting problem and the proton decay problem.
It can provide also
an elegant mechanism to generate
hierarchical 4-dimensional Yukawa couplings, involving
the quasi-localization of matter fields in extra dimension.
We discuss the Kaluza-Klein
threshold corrections to low energy gauge couplings
in generic 5-dimensional orbifold grand unified theories
with quasi-localized matter fields, and apply the results
to some class of $SU(5)$ models on $S^1/Z_2\times Z_2$.
\end{abstract}
\maketitle

\section{Introduction}

Grand unification of the strong and electroweak forces
is a highly persuasive idea for physics at high
energy scales.
However conventional 4-dimensional (4D)
grand unified theories (GUTs) have suffered
from the doublet-triplet splitting
problem and also the problem of too rapid proton decay.
GUTs in (orbifolded)
higher dimensional spacetime can avoid these problems 
by employing the symmetry breaking by 
boundary conditions \cite{kawamura}.
Extra dimension can provide
also an elegant mechanism to 
generate hierarchical Yukawa
couplings \cite{yukawa}.
The quark and lepton fields can be quasi-localized
in extra dimension in a natural manner, and then 
their 4D Yukawa couplings are determined by
the wavefunction factor 
$e^{-M\pi R}$ where
$M$ is a combination of mass parameters in higher dimensional
theory and $R$ is the radius of extra dimension.
This allows that hierarchical Yukawa couplings are obtained from
fundamental mass parameters having the same order of magnitudes.

In any GUT, heavy particle threshold effects at
GUT symmetry breaking scale 
should be taken into account for a precision analysis of
low energy gauge couplings $g_a^2$.
In conventional 4D GUT,  those threshold corrections 
to $1/g_a^2$ are generically of the order of $1/8\pi^2$
and thus {\it not} so important.
However 
higher dimensional field theory 
contain (infinitely) many Kaluza-Klein (KK) 
modes, so can have a sizable threshold correction \cite{choi}.
It is then essential to include the KK threshold
corrections in the analysis 
of low energy couplings in higher dimensional
gauge theories.
In this talk, we discuss the KK threshold corrections
in generic 5D orbifold field theories
with quasi-localized matter fields \cite{choi1}, and apply
the results to some class of $SU(5)$ models on $S^1/Z_2\times Z_2$.

\section{Yukawa Couplings of Quasi-Localized Matter Fields}

Hierarchical Yukawa couplings can be naturally
obtained if the matter fermions are quasi-localized in 
extra dimension \cite{yukawa}.
To see this, consider a 5D theory on $S^1/Z_2$
with coordinate $y\equiv y+2\pi R$,
containing generic bulk fermions and also a brane Higgs field
confined at $y=\pi R$ :
\bea
S=&&-\int d^5x \,\l[ \f{}{}
\,i \bar{\Psi}_I(\gamma^MD_M+
{M}_I\epsilon(y))\Psi_I
\r.\nonumber \\
&&\l.+\delta(y-\pi R)
\l(\, D_\mu HD^\mu H^*+
\f{\lambda_{IJ}}{\Lambda} H\psi_I\psi_J\,\r)\,\r]\,.
\eea
Here $\epsilon (y)=y/|y|$,
$\Lambda$ denotes the cutoff scale, and
$\lambda_{IJ}$ are dimensionless brane Yukawa couplings.
The 5D fermion $\Psi_I$
has the boundary condition
$$
\Psi_I(-y)=z_I\gamma_5\Psi_I(y), \quad
\Psi_I(-y')=z_I \gamma_5\Psi_I(y')\,,
$$
where $y'=y-\pi R$, $z_I=\pm 1$,
and
$\psi_{I}=\f{1}{2}(1+\gamma_5)\Psi_{I}$
(if $z_I=1$) or
$\f{1}{2}(1+\gamma_5)\Psi_{I}^c$
(if $z_I=-1$).
For any value of the kink mass $M_I\epsilon(y)$,
$\Psi_I$ has a chiral zero mode
\beq
\psi_{0I}=\exp (-z_IM_IRy)\,,
\eeq
which is quasi-localized at $y=0$ or $\pi$ depending
upon the sign of $M_I$.
The 4D Yukawa couplings of these quasi-localized 
zero modes are given by
\beq
\label{canonicalyukawa}
y_{IJ} =\sqrt{Z(z_I{M}_I)Z(z_J{M}_J)}\,\,
\lambda_{IJ}
\eeq
where 
$$
Z(M) =\f{M}{\Lambda}\f{1}{e^{2M\pi R}-1}.
$$
Obviously, the 4D Yukawa couplings
$y_{IJ}$ can have hierarchical values
even when  the values of $\lambda_{IJ}$ 
are similar to each other.
For instance, if $z_IM_I$ and $z_JM_J\lesssim - 1/R$,
we have
\beq
y_{IJ}\approx \lambda_{IJ}\sqrt{\l|{M_IM_J}/{\Lambda^2}\r|}
\,,
\eeq
while for $z_IM_I$ and $z_JM_J\gtrsim 1/R$,
\beq
\label{smallyukawa}
y_{IJ}\approx \lambda_{IJ}\sqrt{\l|{M_IM_J}/{\Lambda^2}\r|}\,
e^{-(z_IM_I+z_JM_J)\pi R}\,.
\eeq
The physical interpretation of this result is simple.
If $z_{I,J}M_{I,J}\gtrsim 1/R$, 
the corresponding zero modes are quasi-localized at  $y=0$, 
so the Yukawa couplings are exponentially 
suppressed as they originate from $y=\pi$.
On the other hand, for $z_{I,J}M_{I,J}\lesssim
-1/R$, the zero modes are localized at $y=\pi$, so
there is no suppression of Yukawa couplings.

\section{Kaluza-Klein Threshold Corrections}

The model we study here is a generic 5D gauge theory 
on $S^1/Z_2\times Z_2$ which is described by
\bea
S=-\int d^4 x d y &&\left[ \f{1}{4\hat{g}_{5a}^2} F^{aMN} F^a_{MN}
+D_M \phi D^M \phi^* +\hat{m}^2\phi\phi^*
\r.\nonumber \\ 
&&\quad\quad\l.\,+i\, \bar{\Psi} (\gamma^M D_M +M\epsilon(y)) \Psi \right],
\eea
where
\bea
\label{mass}
&&\f{1}{\hat{g}_{5a}^2}=
\f{1}{g_{5a}^2}+\delta(y)\f{1}{g_{0a}^2}
+\delta(y-\pi R)\f{1}{g_{\pi a}^2}\,,
\nonumber \\
&&\hat{m}^2=m^2 + 2\mu\delta(y) - 2\mu'\delta(y-\pi R).
\nonumber
\eea
The 5D fields in the model can have arbitrary 
$Z_2\times Z_2$ boundary condition,
\bea
&&\phi(-y)=z_\phi \phi(y)\,,\quad
\phi(-y')=z^{\prime}_\phi \phi(y')\,,
\nonumber \\ 
&&\Psi(-y)=z_\Psi \gamma_5\Psi(y)\,,\quad
\Psi(-y')=z^{\prime}_\Psi \gamma_5\Psi(y')\,,
\nonumber  \\
&& A^a_\mu(-y)=z_a A^a_\mu(y)\,,\quad
A^a_\mu(-y')=z^{\prime}_a A^a_\mu(y')\,,
\eea
with $z_\Phi,z^{\prime}_\Phi=\pm 1$ 
($\Phi=\phi,\Psi, A^a_M\,$).
A 5D complex scalar field $\phi$ has a zero mode 
for any value of $R$ if $z_\phi=z_\phi'=1$ and
$m=\mu=\mu'$, while $\Psi$ has a chiral zero mode
for arbitrary  values of $M$ and $R$ if $z_\Psi=z_\Psi'=1$.

The one-loop gauge couplings at low momentum scale $p$
can be written as 
\bea
\label{gaugecoupling}
\f{1}{g_a^2(p)}=\l(\f{1}{g_a^2}\r)_{\rm bare}+
\f{1}{8\pi^2}\l[\Delta_a(\ln\Lambda,m,\mu,\mu',M,R)+
b_a\ln\l(\f{\Lambda}{p}\r)\r]\,,
\eea
where 
$$
\l(\f{1}{g_a^2}\r)_{\rm bare}=\f{\pi R}{g_{5a}^2}
+\f{1}{g_{0a}^2}+\f{1}{g_{\pi a}^2}+
\f{\gamma_a}{24\pi^3}\Lambda\pi R\,,
$$ 
$\Delta_a$ stand for  
KK threshold corrections,
and $b_a$ are the 4D one-loop beta function
coefficients.
Note that $\Delta_a$
are {\it not} sensitive to the unknown physics at $\Lambda$,
so are calculable within orbifold field theory,
while the linearly divergent one-loop corrections in 
$(1/g_a^2)_{\rm bare}$ 
are UV-sensitive and thus {\it not} calculable.
The computation of $\Delta_a$
involves the summation over all massive
KK modes \cite{stephan},  yielding \cite{choi1}
\bea
\label{result2}
\Delta_a =&&
-\f{1}{6}T_a({\phi^{(0)}}_{++})\ln \l(\f{\Lambda(e^{m_{++}\pi R}
-e^{-m_{++} \pi R})}{2m_{++}}\r)
\nonumber \\
&&-\f{1}{6}T_a(\phi_{++})\ln\l(\f{(m_{++}+\mu_{++})(m_{++}-\mu^\pr_{++})
e^{m_{++}\pi R}}{2m_{++}\Lambda}\r.
\nonumber \\
&&\quad\quad\quad\quad\quad\quad\quad
\l.-\frac{(m_{++}-\mu_{++})(m_{++}+\mu^\pr_{++})e^{-m_{++}\pi R}}
{2m_{++}\Lambda}\r)
\nonumber \\
&&-\f{1}{6}T_a(\phi_{+-})\ln\l(
\f{(m_{+-}+\mu_{+-})e^{m_{+-}\pi R}+(m_{+-}-\mu_{+-})e^{-m_{+-}\pi R}}
{2m_{+-}}\r)
\nonumber \\
&&-\f{1}{6}T_a(\phi_{-+})\ln\l(
\f{(m_{-+}-\mu^\pr_{-+})e^{m_{-+}\pi R}+(m_{-+}+\mu^\pr_{-+})e^{-m_{-+}\pi R}}
{2m_{-+}}\r)
\nonumber \\
&&-\f{1}{6}T_a(\phi_{--})\ln\l(
\f{\Lambda (e^{m_{--}\pi R}-e^{-m_{--}\pi R})}{2m_{--}}\r)
\nonumber \\
&&-\f{2}{3}T_a(\Psi_{++})\ln\l(\f{\Lambda(e^{{M}_{++}\pi R}
-e^{-{M}_{++}\pi R})}{2{M}_{++}}\r)
\nonumber \\
&&-\f{2}{3}T_a(\Psi_{+-})\ln \l(e^{-{M}_{+-}\pi R}\r)
\nonumber \\
&&-
\f{2}{3}T_a(\Psi_{-+})\ln\l(e^{{M}_{-+}\pi R}\r)
\nonumber \\
&&-\f{2}{3}T_a(\Psi_{--})\ln\l(
\f{\Lambda (e^{{M}_{--}\pi R}-e^{-{M}_{--}\pi R})}
{2{M}_{--}}\r)\,
\nonumber \\
&&+\frac{21}{12}\l[T_a(A^M_{++})+T_a(A^M_{--})\r]\ln(\Lambda\pi R)
\eea
where the subscripts 
represent the $Z_2\times Z_2'$ boundary conditions.
Here ${\phi^{(0)}}_{++}$ is a 5D scalar field
having zero mode, i.e. scalar field with
$\mu=\mu^\pr=m$, and $\phi_{\pm\pm}$ stand for other scalar fields
{\it without} zero mode.
The one-loop beta function coefficients $b_a$ are given by
\beq
b_a=-\f{11}{3}T_a(A^M_{++})+\f{1}{6}T_a(A^M_{--})
+\f{1}{3}T_a({\phi}^{(0)}_{++})+\f{2}{3}T_a(\Psi_{++})
+\f{2}{3}T_a(\Psi_{--})
\nonumber
\eeq
since
$A^M_{++}$ gives a massless 4D vector, $A^M_{--}$
a massless {\it real} 4D scalar, and $\Psi_{\pm\pm}$
a massless 4D chiral fermion.
In supersymmetric limit, we have 
\bea
\label{result3}
\l(\Delta_a\r)_{\rm SUSY} && = \l[ T_a({V}_{++})
+ T_a({V}_{--})\r]\ln (\Lambda \pi R)
\nonumber \\
&&-T_a({H}_{++})\ln \l(\f{\Lambda(e^{{M}_{++}\pi R}
-e^{-{M}_{++}\pi R})}{2{M}_{++}}\r)
\nonumber \\
&&-T_a({H}_{+-})\ln \l(e^{-{M}_{+-}\pi R}\r)
\nonumber \\
&&-T_a({H}_{-+})\ln \l(e^{{M}_{-+}\pi R}\r)
\nonumber \\
&&-T_a({H}_{--})\ln \l(\f{\Lambda(e^{{M}_{--}\pi R}
-e^{-{M}_{--}\pi R})}{2{M}_{--}}\r)
\eea
where $V_{zz'}$ denotes a vector multiplet,
and $H_{zz'}$ is a hypermultiplet with kink mass
$M_{zz'}$. 
In fact, $(\Delta_a)_{\rm SUSY}$ 
can be computed by a different method 
based on  4D effective supergravity \cite{choi2},
which gives the same result as the direct summation
over the KK modes.
A similar calculation of
KK threshold effects  can be done for 5D theories with 
warped extra dimension \cite{choi2}.

\section{Application to Orbifold GUTs}

To see the importance of
KK threshold corrections,
let us consider a class of 
5D $SU(5)$ orbifold GUTs whose effective 4D theory 
is the MSSM.
To break $SU(5)$ by orbifolding,
$Z_2\times Z_2^\pr$ is embedded into $SU(5)$ as 
\bea
{Z_2} = {\rm
    diag}(+1,+1,+1,+1,+1)\,, \nonumber \\
Z^\pr_2 = {\rm
    diag}(+1,+1,+1,-1,-1)\,. 
\eea
The model contains matter hypermultiplets 
$F_p(\bar{5})$, $F^\pr_p(\bar{5})$, $T_p(10)$ and 
$T^\pr_p(10)$ ($p=1,2,3$) 
with kink masses
${M}_{F_p}$, ${M}_{F^\pr_p}$, 
${M}_{T_p}$ and ${M}_{T^\pr_p}$, 
and also the Higgs hypermultiplets $H(5)$ and $H^\pr(\bar{5})$
with kink masses 
${M}_H$ and
${M}_{H^\pr}$,
where the numbers in brackets represent the $SU(5)$ representation.
We assign the $Z_2\times Z_2^\pr$ parities of these hypermultiplets
as
\bea
\label{bc2}
&&Z_2(F_p)=Z_2(F^\pr_p)=Z_2(T_p)=Z_2(T^\pr_p)=Z_2(H)=Z_2(H^\pr)=1\,,
\nonumber \\
&& Z_2^\pr(F_p)=-Z_2^\pr(F^\pr_p)=
Z^\pr_2(T_p)=-Z^\pr_2(T^\pr_p)=Z^\pr_2(H)=Z_2^\pr(H^\pr)
=-1\,.
\nonumber
\eea
Then using (\ref{result2}), we find
that the QCD coupling constant
at the weak scale is predicted to be \cite{choi1}
\bea
\label{qcd}
\f{1}{\alpha_3 (M_Z)} 
 = 7.8+\f{1}{2\pi}\l[\,\Delta_{\rm gauge}
+\Delta_{\rm higgs}+\Delta_{\rm matter}\r]+{O}\l(
\f{1}{\pi}\r)
\eea
where  
\bea
\f{1}{2\pi}\Delta_{\rm gauge}=\f{3}{7\pi}\,\ln (\pi R\Lambda)
\approx 0.8\nonumber
\eea
corresponds to the KK threshold correction from
the 5D vector multiplet,
\bea
\label{higgs}
\Delta_{\rm higgs}=\f{9}{14}\, \l[\,\ln \l(
\f{\sinh \,\pi R{M}_{H}}{\pi R {M}_{H}}
\f{\sinh \,\pi R {M}_{H^\pr}}{\pi R M_{H'}}\r)+ 
\pi R(M_H+{M}_{H^\pr})  \right]
\nonumber
\eea
is the correction from Higgs hypermultiplets, and
\bea
\label{matter}
\Delta_{\rm matter}&=&
\f{9}{14} \sum_p\l[\,\ln \l(\f{\sinh \,\pi R{M}_{F_p}}
{\pi R{M}_{F_p}}
\f{\pi RM_{F'_p}}{\sinh \,\pi R{M}_{F^\pr_p}}\r)
+\pi R(M_{F_p}-{M}_{F^\pr_p})\,\r]
\nonumber \\
&+&\f{3}{2} \sum_p \l[\,
\ln\l(\f{\sinh \, \pi R{M}_{T_p}}{\pi R{M}_{T_p}}
\f{\pi RM_{T'_p}}{\sinh \, \pi R {M}_{T^\pr_p}}\r) 
+\pi R(M_{T_p}-{M}_{T^\pr_p})\,\r]
\nonumber
\eea 
is the correction from matter hypermultiplets.
For the prediction (\ref{qcd}), we made the usual
assumption \cite{strong} that the theory is strongly coupled at
$\Lambda$, so 
the brane gauge couplings at $\Lambda$
are estimated as $1/g_{0a,\pi a}^2=O(1/8\pi^2)$. 

In order to be consistent with $(1/\alpha_3(M_Z))_{\rm exp}=8.55\pm 0.15$,
$\Delta_{\rm higgs}+\Delta_{\rm matter}$ are required to
be {\it not} so large, which is a nontrivial condition
for the model.
A simple way to avoid a too large $\Delta_{\rm higgs}$
is that both Higgs hypermultiplets
have  ${M}\pi R\ll -1$.
Then the Higgs zero modes are localized at $y=\pi R$
and
\beq
\frac{1}{2\pi}\Delta_{\rm higgs}\approx-\f{9}{14\pi}\,\ln\l(
\sqrt{{M}_{H}{M}_{H^\pr}}\,\pi R\r)
\approx -0.45\,,
\eeq
where the final number is obtained for
${M}_H\pi R\approx {M}_{H^\pr} \pi R\approx -10$.
For matter hypermultiplets,
to generate hierarchical 4D Yukawa couplings, 
some kink masses should be
positive, while some others are negative.
If the Higgs zero modes are localized at $y=\pi R$,
the matter hypermultiplets with $M\pi R\gg 1$ 
have small Yukawa couplings $Y$ suppressed by $e^{-M\pi R}$.
The above form of $\Delta_{\rm matter}$
shows that such hypermultiplets give a contribution
of $O(M\pi R)=O(\ln Y)$ to $\Delta_{\rm matter}$.  
Then, in order for $\Delta_{\rm matter}$ to be small enough,
one needs a {\it nontrivial cancellation} between the contributions
from different matter hypermultiplets.
A simple example realizing such cancellation
is
\beq
\label{su2}
{M}_{F_p}={M}_{F^\pr_p}\,,
\quad 
{M}_{T_p}={M}_{T^\pr_p}\,,
\eeq
for which  $\Delta_{\rm matter}= 0$.

\section{Conclusion}

In this talk, we discussed the KK threshold corrections
to low energy gauge couplings $g_a^2$ from bulk matter fields
whose zero modes are quasi-localized to generate
hierarchical 4D Yukawa couplings. 
We presented the explicit form of threshold corrections
in generic 5D orbifold field theory on $S^1/Z_2\times
Z_2^\pr$.
The typical size of KK threshold corrections to
$1/g_a^2$ is of the order of
$\ln (Y)/8\pi^2$ where $Y$ denotes 
small 4D Yukawa couplings generated by
quasi-localization.
So  KK threshold corrections
can significantly affect the
gauge coupling unification. 
We applied these results to some class of 5D $SU(5)$
models to find the condition to 
get hierarchical Yukawa couplings without spoiling
the successful gauge coupling unification.

\section*{Acknowledgments}
This work was supported in part by
KRF PBRG 2002-070-C00022.



\begin{thebibliography}{0}





\bibitem{kawamura}
Y.~Kawamura, Prog. \ Theor. \ Phys. \ {\bf 105}, 691 (2001);
Prog.\ Theor.\ Phys.\  {\bf 105}, 999 (2001);
G.~Altarelli and F.~Feruglio,
Phys.\ Lett.\ {\bf B511}, 257 (2001);
L.~J.~Hall and Y.~Nomura,
Phys.\ Rev.\ {\bf D64}, 055003 (2001);
Phys.\ Rev.\ {\bf D65}, 125012 (2002);
Phys.\ Rev.\ {\bf D66}, 075004 (2002);
A.~Hebecker and J.~March-Russell,
Nucl.\ Phys.\ {\bf B613}, 3 (2001);
Nucl.\ Phys.\ {\bf B625}, 128 (2002);
T.~Asaka, W.~Buchmuller and L.~Covi,
Phys.\ Lett.\ B {\bf 523}, 199 (2001);
R. Dermisek and A. Mafi, Phy. Rev. {\bf D65}, 055002 (2002);
H.~D.~Kim, J.~E.~Kim and H.~M.~Lee,
JHEP {\bf 0206}, 048 (2002);
H.~D.~Kim and S.~Raby,
JHEP {\bf 0301}, 056 (2003).



\bibitem{yukawa}
N.~Arkani-Hamed and M.~Schmaltz,
Phys.\ Rev.\ {\bf D61}, 033005 (2000);
E.~A.~Mirabelli and M.~Schmaltz,
Phys.\ Rev.\ {\bf D61}, 113011 (2000);
G.~R.~Dvali and M.~A.~Shifman,
Phys.\ Lett.\ B {\bf 475}, 295 (2000);
D.~E.~Kaplan and T.~M.~Tait,
JHEP {\bf 0006}, 020 (2000);
JHEP {\bf 0111}, 051 (2001);
M.~Kakizaki and M.~Yamaguchi,
hep-ph/0110266;
S.~J.~Huber and Q.~Shafi,
Phys. Lett. {\bf B498}, 256 (2001);
A.~Hebecker and J.~March-Russell,
Phys.\ Lett. {\bf B541}, 338 (2002);
F. del Aguila and J. Santiago, JHEP {\bf 0203}, 010 (2002);
Y.~Grossman and G.~Perez,
Phys.\ Rev.\ D {\bf 67}, 015011 (2003);
R.~Kitano and T.~j.~Li,
Phys.\ Rev.\ D {\bf 67}, 116004 (2003);
C. Biggio, F. Feruglio, I. Masina and M. Perez-Victoria,
hep-ph/0305129;
K.~Choi, D.~Y.~Kim, I.~W.~Kim and T.~Kobayashi,
hep-ph/0305024.



\bibitem{choi}
K.~Choi,
Phys.\ Rev.\  {\bf D37}, 1564 (1988);
V.~S.~Kaplunovsky,
Nucl.\ Phys.\  {\bf B307}, 145 (1988)
[Erratum-ibid.\ B {\bf 382}, 436 (1992)].




\bibitem{choi1}
K. Choi, I.-W. Kim and W. Y. Song, hep-ph/0307365.



\bibitem{stephan}
S.~Groot Nibbelink,
Nucl.\ Phys.\ {\bf B619}, 373 (2001);
R.~Contino and A.~Gambassi,
J.\ Math.\ Phys.\  {\bf 44}, 570 (2003).

\bibitem{choi2}
K.~Choi, H.~D.~Kim and I.~W.~Kim,
JHEP {\bf 0211}, 033 (2002);
JHEP {\bf 0303}, 034 (2003);
K.~Choi and I.~W.~Kim,
Phys.\ Rev.\ {\bf D67}, 045005 (2003).

\bibitem{strong}
Z.~Chacko, M.~A.~Luty and E.~Ponton,
JHEP {\bf 0007}, 036 (2000);
Y.~Nomura,
Phys.\ Rev.\ D {\bf 65}, 085036 (2002).











\end{thebibliography}
\end{document}